\documentclass[useAMS,usenatbib]{mn2e}

\usepackage{graphicx}

\title[Observations of microglitches in HartRAO radio pulsars]{Observations 
of microglitches in HartRAO radio pulsars}
\author[A. E. Chukwude and J. O. Urama]{A. E. Chukwude$^{1,2}$\thanks{E-mail:
aus$_{-}$chukwude@yahoo.com}, and J. O. Urama$^{1}$ \\
$^{1}$Department of Physics \& Astronomy, University of Nigeria, Nsukka, Enugu 
State, Nigeria\\
$^{2}$Federal University of Petroleum Resources (FUPRE), PMB 1221 Effurun, Delta
State, Nigeria\\
}
\begin{document}


\pagerange{\pageref{firstpage}--\pageref{lastpage}} \pubyear{2009}

\maketitle

\label{firstpage}

\begin{abstract}
A detailed observation of microglitch phenomenon in relatively slow radio
pulsars is presented. Our analyses for these small amplitude jumps in pulse
rotation frequency ($\nu$) and/or spin down rate ($\dot{\nu}$) combine the 
traditional manual detection method (which hinges on careful visual inspections 
of the
residuals of pulse phase residuals) and a new, and perhaps more objective,
automated search technique (which exploits the power of the computer, rather than 
the eyes, for resolving discrete events in pulsar spin parameters). The
results of the analyses of a sample of 26 radio pulsars reveal that: (i) only 20
pulsars exhibit significant fluctuations in their arrival times to be considered 
suitable for meaningful microglitch analyses; (ii) a phenomenal 299
microglitch events were identified in $\nu$ and/or $\dot{\nu}$: 266 of 
these events were found to be simultaneously significant in $\nu$ and $\dot{\nu}$, 
while 19 and 14 were noticeable only in $\nu$ and $\dot{\nu}$, respectively; 
(iii) irrespective of sign, the microglitches have fractional sizes which cover
about 3 orders of magnitude in $\nu$ and $\dot{\nu}$ ($10^{-11} < 
|\Delta{\nu}/\nu| < 2.0\times10^{-8}$ and $5.0\times10^{-5} < 
|\Delta{\dot{\nu}}/\dot{\nu}| < 2.0\times10^{-2}$) with median values as
$0.78\times10^{-9}$ and $0.36\times10^{-3}$, respectively. 
 
\end{abstract}


\section{INTRODUCTION}

Discrete discontinuities in the rotation of pulsars are broadly of two sorts:
macroglitches and microglitches (e.g.\ Cordes, Downs \& Krause-Polstorff 1988).
Macroglitches (conventionally known as glitches) are characterised by a sudden 
increase in pulsar rotation frequency, $\nu$, usually
accompanied by an increase in magnitude of the spin-down rate, $\dot{\nu}$, 
(e.g.\ Urama 2002; Wang et al.\ 2001; Lyne et al.\ 1996; Flanagan 1990; 
Lyne 1987; Lyne \& Pritchard 1987). Macroglitches are arguably the more 
spectacular
events, characterised by fractional changes in pulsar spin frequency
($\Delta{\nu}/\nu$) and spin-down rate ($\Delta{\dot{\nu}}/\dot{\nu}$) in
the range of $\sim$ 10$^{-8}$ $-$ 10$^{-6}$ and 10$^{-4}$ $-$ 10$^{-1}$,
respectively, (Lyne, Shemar \& Smith 2000; Shemar \& Lyne 1996). To date,
about 280 macroglitches have been reported in about 100, mostly young, pulsars 
(Melatos, Peralta \& Wyithe 2008 and references therein). Macroglitches have  
been extensively studied and are now firmly associated with some unique 
features. The jumps in the rotation frequency ($\Delta{\nu}$) and 
the spin-down rate ($\Delta{\dot{\nu}}$) have a definite signature, 
($\Delta{\nu}$, $\Delta{\dot{\nu}}$) = (+, $-$), corresponding to increases
in the pulsar spin rates and magnitude of the spin-down rates (Melatos et
al.\ 2008 and references therein). It is worthy of note that 
positive $\Delta{\dot{\nu}}$ have been reported in a few macroglitches (with
$\Delta{\nu}/\nu > 10^{-7}$: see, for example, Middleditch et al.\ 2006;
Urama \& Okeke 1999 and references therein). However, we should also mention
that the error estimates in the measured values of 
$\Delta{\dot{\nu}}/\dot{\nu}$ in all these cases are, uncomfortably, too
large. For most well observed macroglitches, the jumps in the rotation 
frequency are accompanied by periods of relaxation towards the unperturbed 
pre-glitch values, which vary on a wide range of timescales (Lyne et al.\ 
1996; Flanagan 1995, 1990; Cordes et al.\ 1988; Lyne 1987; Lyne \& Pritchard 
1987). Macroglitches are now widely believed to originate from some form of 
dynamical change within neutron star interior, such as a sudden and irregular 
transfer of angular momentum from the more rapidly rotating inner superfluid 
components to the slowly spinning crust (Ruderman, Zhu \& Chen 1998; Alpar 
et al.\ 1996, 1984)

On the other hand, microglitches constitute a class of small amplitude, but 
resolvable jumps in both, or either of the pulsar rotation frequency and its 
first time derivative. Unlike the spectacular macroglitch events, the study of 
microglitches have received a far less attention in the last three decades of pulsar 
timing observations. As a result, very little is known about this class of rotational
discontinuities (Chukwude 2002; Cordes et al.\ 1988). To date, only about 90 
microglitches have been reported in $\sim$ 20 radio pulsars (Cordes \& Downs 1985, 
hereafter CD85; Cordes et al.\ 1988; D'Alessandro et al.\ 1995, hereafter DA95). 
The amplitudes of the jumps vary over a wide range, but are generally
believed to be within $\Delta{\nu}/\nu < 10^{-10}$ and $\Delta{\dot{\nu}}/\dot{\nu} 
< 10^{-3}$. However, unlike macroglitches, where the jumps are associated
with a definite signature, microglitches have all possible combination of 
signs and are extraordinarily difficult to understand in the framework of all 
known macroglitch theories (D'Alessandro 1997; Cordes et al.\ 1988; CD85). 
 
Previous observations of microglitches in radio pulsars have relied solely
on visual inspection of carefully calculated residuals of pulse rotation
frequency and its first time derivative ($\Delta{\nu}$ and $\Delta{\dot{\nu}}$, 
respectively). However, observational constraints have permitted only indirect 
calculations of $\Delta{\nu}$ and $\Delta{\dot{\nu}}$ via successive numerical 
differentiation of phase residuals obtained from 2nd-order model fits to 
barycentric times of pulse arrival (DA95 Cordes et al.\ 1988; CD85). In principle,
scatters in the differentiated phase residuals that are significantly in excess 
of measurement uncertainties are attributed to discrete rotational discontinuities. 
The effectiveness of this technique for microglitch observation has, hitherto,
been plagued with sparsely sampled timing data. This results in a paucity of
data, which introduces significant ambiguities in both the epoch and amplitude 
of candidate events. For instance, the process of numerically differentiating
pulse phase residuals could require data segments with widely varying
lengths. Hence, candidate microglitch events whose epochs are located deep within
the segments could easily be averaged out during the process, thereby reducing 
the overall sensitivity of the technique for microglitch observation.  
 
A more comprehensive observational description of the phenomenon of microglitches 
has, however, become an indispensible part of the current quest for improved 
statistics of radio pulsar microglitches. This is particularly important for the 
much sought-after better insight into the incidence and other features  
of these small amplitude discontinuities in pulsar rotation. Specifically, 
such a description is required in order to ascertain the applicability of
the various models involving the neutron 
star interior and/or magnetospheric torque fluctuations (e.g.\ Glampedakis
\& Andersson 2009; Melatos \& Warszawski 2009; Warszawski \& Melatos 2008;
Melatos \& Peralta 2007; Peralta et al.\ 2006; Link, Epstein \& Baym 1993; 
Alpar et al.\ 1984; Cordes \& Greenstein 1981; Arons 1981). Some potential 
benefits of this include a better understanding of the possible origin of 
microglitches and, perhaps, the relationship, if any, between macroglitches 
and microglitches. 

In this paper, we present an in-depth observation of the phenomenon of
microglitches in a sample of 26 radio pulsars. Our data, which span about 
16 years for each object, are characterised by much improved sampling 
intervals of $\sim$ 0.05 to  40 d. The short observation intervals and the 
resultant high data density allowed us to introduce alternative and, perhaps, 
more objective methods of resolving small amplitude discrete rotational 
discontinuities in radio pulsars. {\bf In a forthcoming paper, we will
compare the microglitching rates and size distributions with the glitch
results recently published by Melatos et al.\ (2008).}  

\section[]{OBSERVATIONS}
Regular timing observations of all objects in the current sample of  26
radio pulsars commenced at Hartebeesthoek Radio Astronomy Observatory 
between 1984 January and 1987 May and are still ongoing. However, only the timing
data accumulated up to 2002 October are reported in this paper. HartRAO
regular pulsar timing observation was, however, interrupted between 1999 June 
and 2000 August due to a major hardware upgrade. Only the pulsars B0833$-$45 and 
B1641$-$45, on real time glitch monitoring program, were scheduled during this 
period. The two pulsars are, nontheless, excluded from current analysis. In HartRAO, 
pulse times of arrival (TOAs) were
measured regularly at intervals $\sim$ 1 $-$ 14 days using the observatory 
26-m parabolic radio telescope. Pulses from radio pulsars were recorded by a 
single 10 MHz bandwidth receiver centred near 13  or 18 cm and no
pre-detection dedispersion hardware was implemented during the period under
consideration.

For each pulsar, detected pulses are smoothed with an appropriate filter-time 
constant, and folded over $N_{\mathrm{p}}$ consecutive rotation periods to beat
down the background noise.  $N_{\mathrm{p}}$ is different for different 
pulsars, but it generally varies between 500 and 5000 for the current sample.
This corresponds to an integration time in the range of $\sim$ 48 s $-$ 32 mins. 
An integration was usually started at a particular second by synchronization 
to the station clock, which is derived from a hydrogen maser and is referenced 
to the Universal Coordinated Time (UTC) via the Global Position satellite (GPS). 
On average, three such on-line integrations were made for each pulsar 
during a typical observing session. Details of data acquisition and reduction
at HartRAO have been described elsewhere (Flanagan 1995).

\section[]{TIMING ANALYSIS AND RESULTS}
The resulting topocentric arrival times were transformed to infinite
observing frequency at the Solar System Barycentre (SSB) using the
Jet Propulsion Laboratory DE200 solar system ephemeris (Standish 1982) and
TEMPO2 software package (Hobbs, Edwards \& Manchester 2006).
Subsequent  modelling of the barycentric times of pulse arrival (hereafter 
referred to as BTOAs) and further analyses were accomplished with the HartRAO 
in-house pulsar timing analysis software (CPHAS), which is based on the standard 
pulsar timing technique (e.g.\ Manchester \& Taylor 1977) and is well described 
in Flanagan (1995). To account for the pulsar deterministic spindown, the BTOAs 
were modelled with a simple Taylor series of the form (e.g.\ Lorimer \&
Kramer 2005)
\begin{equation}
\phi(t) = \phi_{0} + \nu(t-t_{0}) + \frac{1}{2}\dot{\nu}(t-t_{0})^{2},
\end{equation}
where $\phi_{\mathrm{0}}$ is the phase at an arbitrary time 
$t_{\mathrm{0}}$, $\nu$ and $\dot{\nu}$ are, respectively, the rotation frequency
and its first time derivative. Equation~(1) presumes that pulsar deterministic
spindown follows a simple power-law relation of the form $\dot{\nu} =
K\nu^{n}$: where $K$ is a positive constant which depends on neutron star magnetic
moment and moment of inertia and $n$ is the torque braking index. For the 
simplest standard model, in which the torque braking processes are largely
due to vacuum magnetodipole radiation at the pulsar rotation frequency 
(e.g.\ Goldreich \& Julian 1969; Pacini 1968) $n =
3$. The difference between the observed BTOAs and the 
predictions of the best-fit model for pulsar rotation (the phase or, more
appropriately, the time residuals) are shown in Fig.~1 for a selection of six
HartRAO pulsars. The observed phase residuals of six pulsars (B0450$-$18, 
B1133+16, B1426$-$66, B1451$-$68,  B1933+16 and B2045$-$16) are dominated 
by intrinsic scatters of unusual large amplitudes. {\bf The intrinsic
scatter refers to a combination of flux noise and some other additional noise: 
which are largely due to effects such as pulse-to-pulse phase jitter, which 
is intrinsic to the pulsar, and interstellar scintillation, which causes 
fluctuations in the observed pulse strength within the integration time and 
bandwith (Cordes \& Downs 1985). Scatters in BTOA residuals dominated 
by these noise processes more or less have white noise statistics, 
irrespective of their amplitudes (e.g.\ Chukwude 2002).} The sizes of
the BTOAs scatters suggest that they could have completely swallowed up any
genuine, smaller amplitude timing noise activity inherent in these objects.  
For these peculiar pulsars, the rms timing noise ($\sigma_{\mathrm{TN}} = 
\sqrt{\sigma_{\mathrm{R}}^{2} - \sigma_{\mathrm{W}}^{2}}$: where $\sigma_{
\mathrm{R}}$ is the root-mean-squares phase residuals from 2nd-order 
polynomial (i.e.\ a model[$\nu$, $\dot{\nu}$]) fit to the entire data span and 
$\sigma_{\mathrm{W}}$ is the rms 
white noise) is less than 5 mP.  Estimates of $\sigma_{\mathrm{W}}$ were
obtained using pairs of BTOA residual data, from 2nd-order models, whose 
separations do not exceed 1 day (see, e.g.\ Chukwude 2002). The signal-to-noise 
ratio (SNR = $\sigma_{\mathrm{R}} / \sigma_{\mathrm{R}}$) of these objects are 
typically less than 3 and they are excluded from further microglitch 
analysis in this paper. For the remainder of 20 pulsars, the observed SNR
lie between $\sim$ 8 and 700, suggesting a wide dispersion in the level of the 
observed timing activity. The phase residuals of some of the objects (e.g.\ 
B0740$-$28, B1323$-$62, B1356$-$60 and B1749$-$28) are characterised by series of 
sudden slope changes. Such discontinuous rotational behaviour have been widely 
attributed to enhanced activity of small-amplitude discrete rotational jumps 
(Chukwude 2002; D'Alessandro \& McCulloch 1997; CD85).

\begin{figure}
\includegraphics[width=77mm]{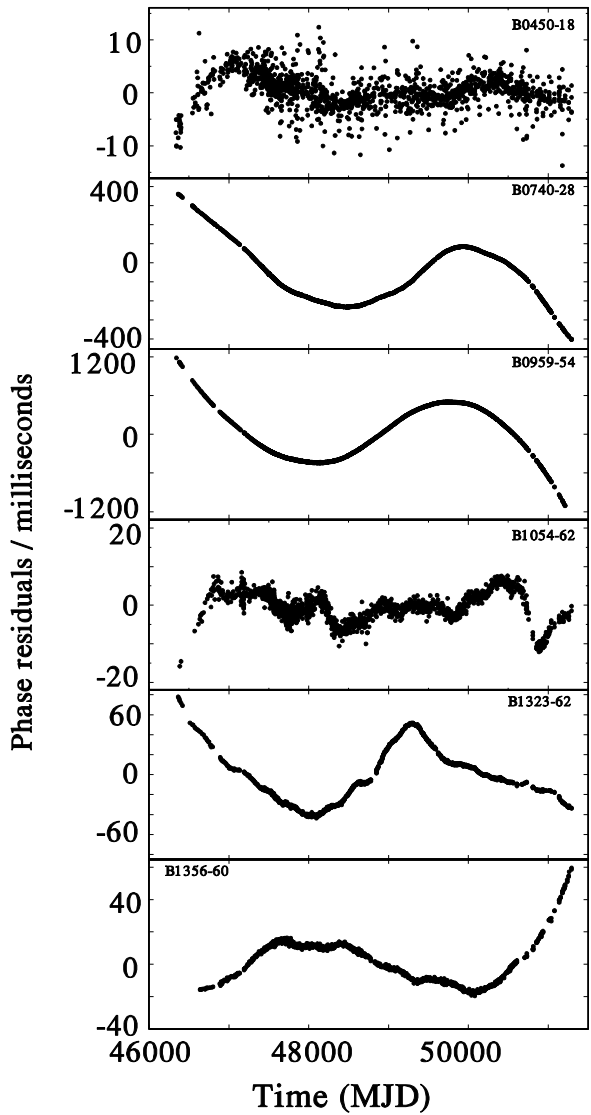}
\caption{The phase residuals, defined in the sense of model-predicted minus
observed BTOAs, for a selection of 6 HartRAO radio pulsars over the interval 
between 1984 and 1999. The pulsar B0450$-$18 typifies objects whose BTOA residuals 
over the $\sim$ 13-yr period are dominated by large amplitude scatter, which have
predominantly white noise statistics. Meaningful microglitch analyses were,
generally, not possible for this category of pulsars using current techniques.
}
\end{figure}

\subsection{BTOA Error Assignment}
Assignment of correct errors to the pulse times of arrival data is crucial
for a meaningful microglitch analysis. Such analysis usually involves
a careful calculation of numerical values of  small amplitude jumps in 
pulsar spin parameters, the associated formal errors and the determination of
the significance of the jump parameters relative to the formal errors. 
Therefore, an accurate BTOA error assignment improves both the sensitivity 
of the technique and the reliability of the results. In order words,
under- or over-estimation of the errors assigned to BTOAs  would seriously
affect the results of any glitch analysis, especially the microglitches. 
Correct assignment of BTOA errors is particularly important 
for the newly introduced automated microglitch detection technique. This 
technique employs the goodness of fit parameter, which relies heavily on
correctly weighted least-square fits, as a tracer of microglitch events. 

Unfortunately, the formal BTOA errors returned by the HartRAO in-house routines 
used to extract arrival times from observations are known to be grossly 
underestimated (Flanagan 1995). Consequently, realistic BTOA errors are obtained 
from the real scatter in the BTOA data (Flanagan 1995). Short segments of data, 
spanning between $\sim$ 50 and 200 days, are modelled with 2nd-order polynomial 
plus a dispersion measure (DM) term. The length of the segments, which depended 
on both the timing activity level and the density of BTOAs, is such that the 
resulting phase resdiduals are statistically equal to zero. Hence, the segment lengths are expected to
be shorter for pulsars with enhanced microglitch activity or near epochs of
macroglitches to compensate for the expected additional scatter. The residuals 
are manually examined for possible outliers, which could bias the error
estimates. Finally, BTOA 
errors are calculated from the real scatter in the phase residuals. Our 
estimator (following Flanagan 1995) is the two-sample variance calculated over 
successive blocks of length $\Delta T$: where $\Delta T \sim$ 5 $-$ 10 days, 
depending on the data sampling rates. The error assigned to each BTOA is the 
rms of the two-sample variance of the block within which it falls. Although this
method, unfortunately, results in BTOA data errors not being strictly
independent, this problem is somewhat alleviated since our model fits
involve data lengths $T$ (where $T \gg \Delta T$). The 13 and 18 cm 
observations were separately analysed, since the
amplitudes of the observed scatter are different at the two frequencies. BTOA
errors calculated using this method are found to be larger than those 
returned by the fitting routine, on average, by factors of $\sim$ 2 and 4 
for the 18 and 13 cm data, respectively. It has been shown (Chukwude 2007; Urama
2002; Flanagan 1995) that BTOA uncertainties obtained in this manner
represent more realistic estimates.

\begin{figure}
\includegraphics[width=75mm]{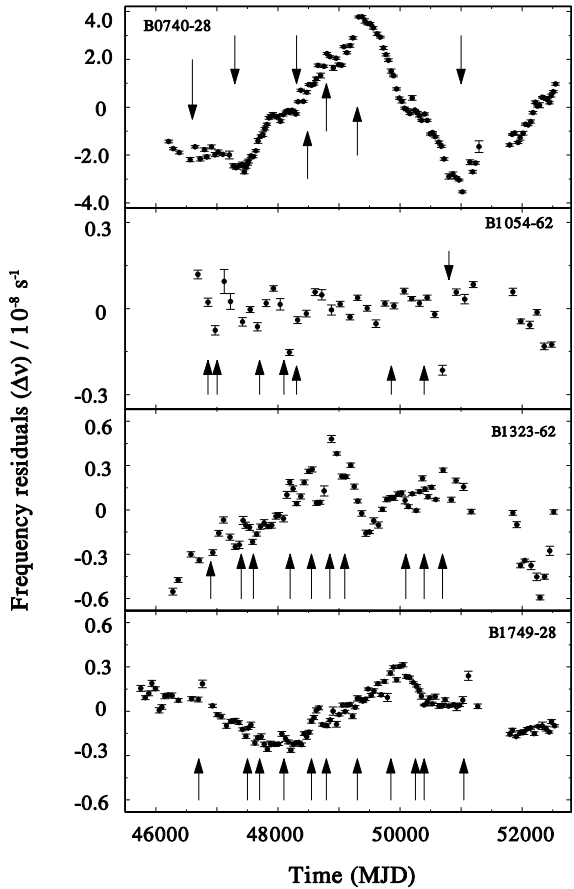}
\caption{Time evolution of the frequency residuals ($\Delta{\nu}(t)$),
calculated as described in the text, for a selection of 4 HartRAO pulsars
over the period between 1984 and 2001. Arrows indicate the points where
scatters in $\Delta{\nu}$ are signficantly ($>$ 2-$\sigma$) larger than the
measurement uncertainties, suggestive of the occurence of real jump in the
pulsar rotation. Error bars are 1-$\sigma$ formal standard errors.}
\end{figure}

\subsection{Manual search for microglitches}
As usual, the manual technique for microglitch detection relies largely on 
visual inspections of plots of carefully calculated residuals of pulsar spin 
frequency ($\Delta{\nu}$) and/or its first derivative ($\Delta{\dot{\nu}}$) for 
scatters significantly in excess of the measurement uncertainties. The frequency 
and spin-down rate residuals can be calculated either directly, using coefficients 
of local fits of 2nd-order polynomial model to short spans of BTOAs (Flanagan 
1995; Cordes et al.\ 1988), or indirectly, by numerically differentiating 
phase residuals (e.g.\ DA95; CD85). The apparent high quality of the current 
data (high density of BTOAs) allowed us to implement the former method, which 
promises to be  more sensitive to small discrete discontinuities. 

For each of the 20 pulsars, a time series of the pulsar rotation frequency 
($\nu(t)$) were calculated by performing weighted least-square fits of a 
2nd-order polynomial to short independent blocks of BTOAs. The block lengths 
were determined by the data sampling rates, the intrinsic scatter in the BTOAs 
and our desire to keep the formal errors in $\nu$ to $\sim$ 1 part in 10$^{9}$. 
These constraints resulted in segment lengths in the range of 35 $-$ 120 days. 
The frequency residuals ($\Delta{\nu}(t)$) were calculated as the 
differences between the raw data ($\nu(t)$: obtained from local fits to 
independent shorts segments of BTOAs) and a model[$\nu$, $\dot{\nu}$] fitted to 
the entire data span length (with epoch near the midpoint of the data). The 
frequency residuals of 4, out of the 20, pulsars used in current analysis are 
shown in Fig.~2. The $\Delta{\nu} - t$ plots were carefully inspected for
evidence for possible discontinuous rotation, usually depicted by scatters 
whose amplitudes are significantly ($>$ 1-$\sigma$) larger than the error bars. 
Generally, such large rotational discontinuities were interpreted as signatures 
of microglitch events and the unrefined epochs of the events were read directly 
from the horizontal axis of the graph. The unrefined epochs of some of the 
candidate microglitch events are marked with arrows in Fig.~2.
 
Once a candidate event had been identified in this manner, short data segments 
spanning $\sim$ 200$-$400-d, and bracketing the candidate microglitch event
were modelled with a 2nd-order polynomial. The resulting phase residuals were 
carefully re-examined for evidence for a discontinuous rotation, easily seen
as a sudden slope change in the observed residuals. Such a sudden slope change 
is attributed to a microglitch event and the epoch is now obtained
directly from the phase residual plots. Fig.~3 shows typical examples of the 
phase residuals from short segments known to habour one or more candidate 
events, first identified with the $\Delta{\nu}$ plots. {\bf Basically, 
possible candidate events are simply depicted by sudden slope changes in 
the observed phase residuals} In principle, this 
approach, generally, results in improved epochs of candidates events, with 
epoch uncertainties reduced to observation intervals.   

The relevant jump parameters (the epoch, the sizes of the jumps in $\nu$ and 
$\dot{\nu}$: $\Delta{\nu}$ and $\Delta{\dot{\nu}}$, respectively)
were estimated by separately modelling the BTOA data extending between 200 
and 400 d on both sides of an identified event with equation~(1). This method 
resulted in two sets of ephemerides for each event. $\Delta{\nu}$ and
$\Delta{\dot{\nu}}$ were estimated, respectively, from $\Delta{\nu} = 
\nu_{\mathrm{post}} - \nu_{\mathrm{pre}}$ and $\Delta{\dot{\nu}} = 
\dot{\nu}_{\mathrm{post}} - \dot{\nu}_{\mathrm{pre}}$ (Chukwude 2002).
The parameters $\nu_{\mathrm{pre}}$ and $\dot{\nu}_{\mathrm{pre}}$
refer to the pre-event values extrapolated to the event epoch, while        
$\nu_{\mathrm{post}}$ and $\dot{\nu}_{\mathrm{post}}$ are the corresponding
post-event values, also extrapolated to the event epoch. Current method 
identified 124 candidate discrete jumps in $\nu$ and/or $\dot{\nu}$: about 
80 were noticeable simultaneously in $\nu$ and $\dot{\nu}$ while 25 and 19 
were identified in only $\nu$ and $\dot{\nu}$, respectively. 

\subsection{Automated search for  microglitches}
The current automated (or at least semi-automated) method for detecting
microglitches was developed and applied in analysis of Vela pulsar data
(Flanagan 1995). The technique employs a statistic, the ``goodness of fit"
parameter ($Q$), to determine when the model fit to BTOAs suddenly becomes 
too bad, probably, due to discrete rotational discontinuity. 
In HartRAO, the least-squares fitting routine employed in pulsar timing
analysis uses a version of CURFIT of Bevington (1969), as modified by
Flanagan (1995). Some of the outputs of the modified CURFIT include the
chi-square ($\chi^{2}$) and the reduced chi-square ($\chi_{R}^{2} = 
\chi^{2}/{dof}$, where $dof$
is the number of degrees of freedom of the fit). The goodness of fit
parameter $Q$ is estimated by calculating the probability of obtaining the
measured value of $\chi_{R}^{2}$ from the data, assuming that the fitted
model is the correct one. Following Flanagan (1995), $Q = Q(dof/2,
\chi^{2}/2)$, where $Q = Q(dof/2, \chi^{2}/2)$ is the incomplete gamma
function. However, the calculation of meaningful $Q$ values is strongly 
dependent on the availability of sensible estimates of the BTOA errors 
employed in the weighted least-squares fits. As a result, extra efforts 
are required in order to estimate realistic pulse times of arrival 
uncertainties provided to the fitting routine. This is extremely important 
in order to minimise the sensitivity of $Q$ values to seeming outlier BTOA data. 

In this method, weighted least-square fits are performed. About 50$-$100-d block  
of BTOAs was modelled with equation~(1) plus a DM term. The block length was 
subsequently incremented in steps of 7 d, correponding roughly to the mean
observation interval, until $Q$ value suddenly dropped below 0.1. A
broad-brush approach is to interprete such a low $Q$ value as suggesting a 
poor description of the data by the model, due to a small discrete jump in the 
pulsar rotation. The resulting ephemerides were written to a file. The 
procedure was repeated, now starting from the epoch of this 
presumed event, until the end of the data set was reached (hereafter referred to 
as forward pass). The entire process $-$ identification of candidate events,
writing of the ephemerides to a file and restarting the model-fits $-$ is
automated. Hence, once the parameters are properly set up, a complete pass 
through the data was achieved, with minimum manual intervention. To ensure
that very small amplitude events were not averaged out, the $\sim$ 50-d start
block was staggered by $\sim$ 20 d, at a time, and the whole process was 
repeated. {\bf The staggered fits improved the sensitivity of the automated
search technique, as it identified a few more microglitch events which
were completely averaged out in the relatively long start blocks.} 
Finally, the process was repeated starting from the end of a data set, and  
going backwards until the start of the data is reached (hereafter referred to 
as backward pass). 

The results of these procedures were sets of ephemeris files containing ephemerides 
extending on both sides of the candidate microglitch events. The epoch of an event 
was taken as the average of the two arrival times bracketing the event and the jump 
parameters, $\Delta{\nu} = \nu_{\mathrm{post}} - \nu_{\mathrm{pre}}$ and $\Delta{\dot{\nu}} =
\dot{\nu}_{\mathrm{post}} - \dot{\nu}_{\mathrm{pre}}$, were calculated as the difference 
between  consecutive ephemerides after extrapolating to the event epoch. In most 
cases, real discrete discontinuities identified in forward pass were also detected 
in the backward pass with nearly coincident epochs and jump parameters of opposite
signs, but of almost equal magnitudes. The automated technique identified 330 
candidate discrete jumps simultaneously in $\nu$ and $\dot{\nu}$.
Additional 80 and 130 events were resolved separately in $\nu$ and $\dot{\nu}$, 
respectively. As expected, the automated technique equally identified all 
the manually observed candidate events, as evidenced by the nearly coincident
epochs and jump parameters of similar size and signs. However, it was observed 
that the parameters of these microglitches were generally better resolved by 
the manual technique. {\bf Currently, we are not aware of any serious
systematic bias in the techniques implemented in the current analysis. 
A possible bias in the automated search method could have resulted from BTOA 
data points with grossly underestimated errors. However, the resulting spurious 
events, which would be characterised by significantly different epochs in the 
forward and backward passes through the data, are easily eliminated during 
manual inspections of the microglitch data.}

\begin{figure*}
\begin{minipage}{120mm}
\includegraphics[width=100mm, angle=270, clip=']{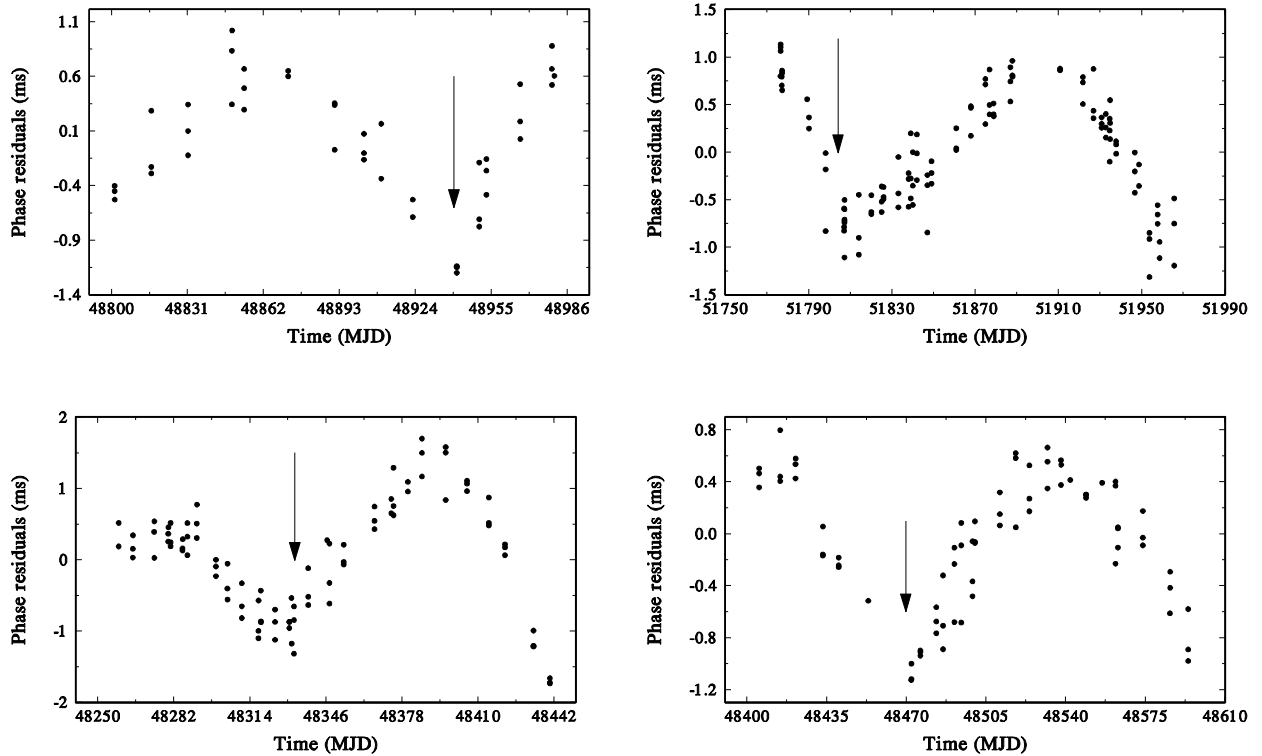}
\caption{The phase residuals around some candidate microglitch events in the
pulsar B0740--28, whose observed timing activity show strong evidence for
enhanced microglitch activity. The arrows indicate epochs sudden slope
changes caused discontinuous rotation}
\end{minipage}
\end{figure*}

\subsection{Significance of the microglitch parameters}
The amplitudes of all identified candidate events in $\nu$ and/or 
$\dot{\nu}$ were tested against the null hypothesis that they are mere
accumulations of random walks of preferred signs. Following  D'Alessandro 
et al.\ (1995), Cordes \& Downs (1985) the significance test was accomplished
by comparing the amplitude of the observed events in $\nu$ and $\dot{\nu}$
with the expected standard deviations of idealized large-rate random walk
process. {\bf This refers to an accumulation of step functions in pulsar 
phase residuals and/or its derivatives whose effects, over a time, 
could mimic genuine discrete rotational discontinuities.} Events are 
considered real, that is their amplitudes are too large to be due to random 
walk fluctuations, 
if $\Delta{\nu} \geq 
\sigma_{\Delta{\nu}}N$ and $\Delta{\dot{\nu}} \geq
\sigma_{\Delta{\dot{\nu}}}N$: where $\sigma_{\Delta{\nu}} = 
\sqrt{S_{1}\Delta{t}}$ and
$\sigma_{\Delta{\dot{\nu}}} = \sqrt{S_{2}\Delta{t}}$ are, respectively, the
standard deviations of idealized random walks in $\nu$ and $\dot{\nu}$ with 
strength parameters $S_{1}$ and $S_{2}$, respectively, $N$ is an integer denoting 
the significance level of the test and $\Delta{t}$ is the difference between
the arrival times bracketing a candidate event. In this paper, $\Delta{t}$ 
represents the upper bound on the rise time of the microglitch events. For the 
current sample, values of $\Delta{t}$ lie between $\sim$ 0.05 and 40 d, with a 
mean value of 10 d.  Several authors (e.g.\ CD85, Cordes et al.\ 1988; DA95) have 
demonstrated that $N = 5$ is an appropriate signficant level, as it correctly 
rejects spurious random walk-induced events which could mimic real
microglitches. Given the similarity between the HartRAO and JPL pulsar data, in 
terms of observing frequencies and BTOA signal-to-noise ratio,  we equally 
adopted $N = 5$ for the current significance test.

Following Chukwude (2002), DA95, CD85, we calculated strength parameters 
($S_1$ and $S_2$, respectively) from BTOAs bracketed by  any two 
identified candidate events (hereafter referred to 
as method A) and from the entire BTOA data available for a pulsar (hereafter
referred to as method B). Our presumption is that data sets used in method A,
probably, contain no discrete discontinuity and hence has extremely low timing 
activity. On the other hand, the data set employed in method B incorporates 
all the rotational discontinuities, including all the microglitches. In 
principle, methods A and B are expected to yield, respectively, the lower 
and upper limiting estimates for the strength parameters. The
bottom line is that, though the strength parameter estimate is variable in
the presence of random walks, its true value should lie within the values 
obtained from the two methods. Since the rms phase residuals of short and
long data span lengths are believed to be dominated by events in $\nu$ and 
$\dot{\nu}$, respectively (DA95; CD85). In view of this fact, we placed
greater weight on methods A and B in testing the significance of $\Delta{\nu}$ 
and $\Delta{\dot{\nu}}$, respectively. In addition, the amplitudes of a 
candidate events were required to exceed the formal standard error by, 
at least, a factor of two. In principle, the necessary condition that must 
be satisfied by the jumps in $\nu$ and $\dot{\nu}$ is that $\Delta{\nu} > 
2\epsilon_{\Delta{\nu}}$ and $\Delta{\dot{\nu}} > 2\epsilon_{\Delta{\dot{\nu}}}$, 
respectively, where $\epsilon_{\Delta{\nu}}$ and $\epsilon_{\Delta{\dot{\nu}}}$ 
are, the formal standard errors in $\Delta{\nu}$ and $\Delta{\dot{\nu}}$,
respectively. The formal standard errors were obtained directly from the HartRAO
in-house timing analysis routine. We believe that these conditions are 
stringent enough to filter out most, if not all, spurious events, which could 
mimic genuine microglitch events.  

Over 600 candidate discrete jumps were identified by a combination
of the manual and automated search techniques for microglitches. However, 
only 299 of these events have amplitudes that are large enough to pass the 
significance tests (and are hereafter referred to as microglitches). The 
remainder of $\sim$ 300 jumps failed the tests and were consequently
rejected, as mere spurious events arising from accumulations of 
large rate steps of random walks. Out of the 299 microglitches, 266 ($\sim$ 90\%) 
are found to be simultaneously significant in both $\nu$ and $\dot{\nu}$ ( 71 of 
these were better resolved by the manual search technique). {\bf These
represent microglitch events that produced detectable changes in the
pulsar spin rate evolution. However, it is believed that several candidate
events might have been averaged out in the process leading to the
calculation of the $\Delta\dot{\nu}$ data. Hence, one expects the more
objective automatic technique to resolve several more microglitches.} 
Others are noticeable in either of the spin parameters ($\nu$ or 
$\dot{\nu}$). The parameters of the 266 and 33 microglitches are summarized 
in Tables~(1) and (2), respectively. A summary of our results shows that 
the total 
microglitches in $\nu$ and $\dot{\nu}$ are 285 and 280, respectively. Of 
the 285 microglitches in $\nu$, 148 and 137 of the events have positive 
and  negative amplitudes, respectively, while 139 and 141 of 
$\Delta{\dot{\nu}}$ are, respectively, positive and negative. The fractional 
jumps in $\nu$ ($\Delta{\nu}/\nu$) have minimum and maximum values as $\sim$ 
$-8\times10^{-9}$ and $2\times10^{-8}$, respectively, while $\sim$ 
$-3\times10^{-3}$ and $4\times10^{-3}$ correspond, respectively, to 
minimum and maximum amplitudes of $\Delta{\dot{\nu}}/\dot{\nu}$. The jumps, 
nonetheless, show no apparent preferential combination of signs ($\Delta{\nu}$, 
$\Delta{\dot{\nu}}$). 

\begin{table*}
\centering
\begin{minipage}{140mm}
\caption{Parameters of the microglitches, whose amplitudes are simultaneously
significant in both $\nu$ and $\dot{\nu}$, for the 20 HartRAO radio pulsars 
observed using both the manual and automatic search techniques.}
\begin{tabular}{@{}llrrrrcllrrrr}
\hline
\multicolumn{1}{l}{Pulsars} & \multicolumn{1}{l}{Epoch} & 
\multicolumn{1}{r}{$\Delta{\nu}/\nu$} & 
\multicolumn{1}{r}{ERR$^{a}$} & \multicolumn{1}{r}{$\Delta{\dot{\nu}}/\dot{\nu}$} 
&  \multicolumn{1}{r}{ERR$^{a}$} &  &   \multicolumn{1}{l}{Pulsars} &
\multicolumn{1}{l}{Epoch} & \multicolumn{1}{r}{$\Delta{\nu}/\nu$} &
\multicolumn{1}{r}{ERR$^{a}$} & \multicolumn{1}{r}{$\Delta{\dot{\nu}}/\dot{\nu}$}
&  \multicolumn{1}{r}{ERR$^{a}$} \\
(PSR B) & (MJD) & ($10^{-9}$) &   & ($10^{-3}$) &  &  & (PSR B) &  (MJD) &
($10^{-9}$) &   & ($10^{-3}$) &  \\
(1)  & (2)  & (3) & (4) & (5) & (6) &  & (1)  & (2)  & (3) & (4) & (5) & (6) \\\hline
0736$-$40&47076&0.43&3&$-$0.15&2& &0835$-$41&46271&$-$0.43&5&2.2&4 \\
 & 47751  &  0.12 &2&$-$0.16&2 & & & 47207  &  0.07&1&0.11&1  \\
 & 47943$^{+}$ &  0.25&3&0.17&2 & & & 47840  &  0.12&1&$-$0.079&3\\
 & 48283  &  0.13&2& 0.75&2& & &49780  & $-$0.13&6&$-$0.064&3\\
 & 48502  & 1.08 & 8 &$-$0.89 & 4& & 0959$-$54& 46479 & 6&2&$-$1.8&5\\
 & 49292$^{+}$ &  0.34&4&0.83 &6& & & 47477$^{+}$ &  2.2&2& 0.32 &6 \\
 & 50395$^{+}$ & $-$0.67&2&$-$0.44&2& & & 48138$^{+}$ &  2.4&4&$-$0.64 &2\\
 & 50612  & 0.17&2&0.42&1 & & & 48490  & $-$1.64&8&$-$0.46&2    \\
 & 51997  & 0.13&2&$-$0.83&3& &  & 48600 &  0.8&2& 0.75 &9\\
 & 52224  & 0.45&6&0.55&8 & & & 49241$^{+}$ & $-$1.79 &8& 0.80 &6 \\\\
 & 52409$^{+}$& $-$0.86&5&$-$1.2&4& & & 50845$^{+}$ &  1.74&8& 1.19 &4\\
0740$-$28&46653$^{+}$&0.46&7&$-$0.052&5& & &52236$^{+}$ & $-$3.7 & 1& 1.98 & 9 \\
 & 46806  &  1.2&4&$-$0.08&2& & 1054$-$62 & 47153 & 0.37&8&$-$0.28&5 \\
 & 46992  & $-$0.17&6&0.057&6& & & 47913  &  0.35&5&0.19&3 \\
 & 47085  & $-$0.06&1&$-$0.026&5& & &48190$^{+}$ &$-$0.95&8& 0.26 &6 \\
 & 47358$^{+}$ & 0.72&3&$-$1.05&9& & & 48301  &  0.69&6&0.18&3 \\
 & 47450  & 0.39&5&$-$0.146&7& & & 48480$^{+}$ & 1.5&2&$-$0.56 &3\\
 & 47576  & $-$0.38&8&$-$0.08&2& &  & 49169  & 0.12&4&0.06&2 \\
 & 47681$^{+}$ & 2.75&8&$-$1.6 & 2& &  & 49378  &  0.29&8&$-$0.19&6\\
 & 47736  & $-$0.55&9&0.09&2& & & 49631  & 0.18&5&$-$0.09&2 \\\\
 & 47843  & $-$0.75&5&$-$0.10&1 & & & 50025  & $-$0.5&2&0.12&3  \\
 & 47956  & $-$0.95&6&0.070&5 & & & 50386  & 0.20&5&0.47&2\\
 & 48084  & $-$0.62&7&0.034&6 & & & 50782$^{+}$ &  3.2&4&$-$0.88&8 \\
 & 48246$^{+}$ & $-$2.3&3&$-$0.99 &2 & & & 50997$^{+}$ &  2.3&4&$-$0.59&5 \\
 & 48345  &  1.4&2&2.5 & 4 & & & 52008  & $-$0.12&3&0.12&3 \\
 & 48427  & $-$0.7 & 2 &$-$0.29 & 4 & & & 52166  & $-$0.75&9&$-$0.11&5 \\
 & 48470$^{+}$&$-$2.4&2& 2.6 &2& & 1221$-$63& 48147 & $-$0.06 &1&$-$0.022&1 \\
 & 48646  &  1.14 & 9 & $-$0.07 &1 & & & 48788$^{+}$ &  0.87&4&0.79 & 7\\
 & 48852$^{+}$  &  1.23  & 8 &0.25&2& & & 49308  &  0.06&2&0.05&1 \\
 & 48940  &  1.2 &2&0.6&2 & & & 49814$^{+}$ &  1.5 &4&$-$0.19&6 \\\\
 & 49050  &  2.8&3 & 1.55 &3 & & & 50830  & $-$0.17&4&0.04&1 \\
 & 49287  &  2.0 &2 & 0.96&3 & & & 50960$^{+}$ & 3.6 &5& 0.96&9 \\
 & 49353$^{+}$ & $-$2.2 & 5 & 0.35 &2 & & & 51931  &  1.2&3&$-$0.9&3 \\
 & 49621& $-$0.16&4&$-$0.092&4& & 1240$-$64& 47460 & 0.18&2&$-$0.12&3\\
 & 49915  &  0.89&4&0.070&3 & & & 47960$^{+}$ & $-$0.33&4&0.09&2 \\
 & 50008  & $-$0.5&2&0.16&2 & & & 48464  & 0.19&2& 0.12&3 \\
 & 50185 &  0.65&8&$-$1.3&2 & & & 49030  & $-$0.033&2& 0.29&2 \\
 & 50376$^{+}$ & 1.19&9& $-$0.9&2 & & & 49610  &  0.32&3&$-$0.13&1 \\
 & 50529  & 0.77&8&$-$0.08&1& & & 50092  & $-$0.68&5&0.18&2 \\
 & 50609&0.4&1&$-$0.18&2& & &50503$^{+}$&$-$0.78&6&$-$0.59&9\\\\
 & 50782  & 1.3&2&1.0&2 & & & 50677  & $-$0.11&3&0.014&2 \\
 & 50978$^{+}$&1.45&8&0.69 &3& & & 51871& $-$0.09&2&$-$0.06&1 \\
 & 51109  & 1.5&2&$-$0.53&3 & & 1323$-$58 & 47586$^{+}$ & 1.43&8&1.5&3\\
 & 51785$^{+}$ & $-$1.93&8&0.96&2& & & 47878  &  $-$0.16&2&1.30&2 \\
 & 51880  & $-$ 0.93 & 8 & 0.38 & 2& & &47955$^{+}$ & $-$1.45&5&1.07&8\\
 & 51970  & $-$0.19&7&$-$0.22&1& & & 48212  & $-$1.03&4&0.85&2\\
 & 52030$^{+}$ & 2.32&9&$-$1.38&9 & & & 48700  & $-$0.69&2&$-$1.28&1 \\
 & 52111  & $-$0.7&2&$-$0.11&3 & & & 49245  & $-$0.78&5&0.43&2 \\
 & 52249  & $-$1.2&2&0.08&2 & & & 49792  & $-$1.82&7&$-$0.82&3\\
 & 52370$^{+}$ & $-$0.28&5&0.25 &2 & & & 49879  &  0.47&4&0.28&2\\\hline
\multicolumn{13}{l}{$^{a}$ The formal standard error and refers to the
 least significant digit}   \\
\multicolumn{13}{l}{$^{+}$ Indicates epochs of events whose jump parameters 
were better resolved by the manual technique}
\\\hline

\end{tabular}
\end{minipage}
\end{table*}

\begin{table*}
\centering
\begin{minipage}{140mm}
\contcaption{}
\begin{tabular}{@{}llrrrrcllrrrr}
\hline
\multicolumn{1}{l}{Pulsars} & \multicolumn{1}{l}{Epoch} & 
\multicolumn{1}{r}{$\Delta{\nu}/\nu$} &
\multicolumn{1}{r}{ERR$^{a}$} &
\multicolumn{1}{r}{$\Delta{\dot{\nu}}/\dot{\nu}$} 
&  \multicolumn{1}{r}{ERR$^{a}$} &  &  \multicolumn{1}{l}{Pulsars} &
\multicolumn{1}{l}{Epoch} & \multicolumn{1}{r}{$\Delta{\nu}/\nu$} &
\multicolumn{1}{r}{ERR$^{a}$} &
\multicolumn{1}{r}{$\Delta{\dot{\nu}}/\dot{\nu}$}
&  \multicolumn{1}{r}{ERR$^{a}$} \\
(PSR B) &  (MJD) & ($10^{-9}$) &   & ($10^{-3}$) &  &  & (PSR B) &  (MJD) &
($10^{-9}$) &   & ($10^{-3}$) &  \\
(1)  & (2)  & (3) & (4) & (5) & (6) &  & (1)  & (2)  & (3) & (4) & (5) & (6)
\\\hline
 & 50150$^{+}$ &  0.99&9&1.23&9& & & 47860 & 1.5 &4 &$-$1.9&5\\
 & 50455  & $-$0.92&6&0.98&3 & & & 48000  &   1.1&3&$-$1.2&3 \\
 & 50670$^{+}$ & $-$0.55&8&$-$0.58&8 & & & 48316& $-$0.99&9&$-$0.32&2 \\
 & 50964 & 0.48 & 8 & $-$0.64 & 5& & & 48548  & 0.75 &6& 0.54 & 5 \\
 & 51020  &  0.5& 1 &$-$0.56&5& & & 49085$^{+}$ &$-$1.9&2&$-$2.46&1 \\
 & 51880  & 1.08 & 7 & $-$0.66 & 4 & & & 49206  &  0.60&3&1.78&9 \\
 & 52338  &$-$1.5 & 3 & $-$1.2 & 4& & & 50210& $-$1.5 & 2 &$-$0.57&3 \\
1323$-$62 & 46769 & $-$0.7&2&0.06&2& & &50673$^{+}$&$-$1.7 &8&$-$2.58&6\\
 & 46940  & $-$1.92&10&$-$0.75&1& & & 52276  &  1.6& 2&1.52&10 \\
 & 47163&$-$1.4 & 3&0.28&4 & & 1449$-$64& 48360&0.22&1&0.042&6 \\\\
 & 47666$^{+}$ &  0.82&6&$-$0.17&1& & & 50212  &   0.05&1&0.014&4 \\
 & 47729$^{+}$ &  $-$0.90&5&0.8&2& & & 50768  &   0.12&5&0.05&2 \\
 & 47875  &   3.8 & 8 &0.6&2 & & & 50835  &  $-$0.08&3&$-$0.06&1 \\
 & 48059  &   0.40&7&$-$0.19&4& & 1556$-$44&48861&$-$0.06&1&0.008&3\\
 & 48242$^{+}$ &  $-$1.3& 4 &0.8&2 & & & 50104  &   0.08&1&$-$0.05&1 \\
 & 48371  &  $-$0.63&8&0.32&3 & & & 51166  &   0.15&5&$-$0.08&1 \\
 & 48459$^{+}$ &   0.67&9&0.48&5& &1557$-$50&47413&$-$0.12&2&$-$0.035&5 \\
 & 48670  & $-$0.56&6&$-$0.9&2& & & 47913  &  $-$0.04&1&0.0130&2 \\
 & 48860$^{+}$ &  0.69&9&0.65&4& & & 47960  &  $-$0.18&2&$-$0.025&4 \\
 & 48939  & $-$1.4 & 2 &0.55&4 & & &48392$^{+}$&$-$0.66&5&$-$0.095&3 \\\\
 & 49150$^{+}$ &  1.6& 2 &0.19&3 & & & 48524  &   0.12&2&0.058&4\\
 & 49334  & $-$1.5 & 3 &$-$0.38&4& & & 48552 &  $-$0.18&4&$-$0.083&6 \\
 & 49408  & $-$0.62&8&$-$0.44&3 & & & 48928$^{+}$ & 0.74&3&$-$0.19&4 \\
 & 49581  &  0.9 & 2 &0.16&4 & & & 49326  &  $-$0.12 &2&0.06 &1 \\
 & 49812  &  0.42&8&$-$0.26&2 & & &49750$^{+}$ & $-$0.28&3&0.092&5  \\
 & 49906  &  $-$0.36&5&0.07&2 & & & 50018  &  $-$0.06&1&0.10&1 \\
 & 50019  &   1.3 & 2 &0.18&3 & & & 50764$^{+}$ &  $-$0.52&2&0.076&5 \\
 & 50104  & $-$0.9& 2 &$-$0.20&4 & & & 51274  &  $-$0.32&5&$-$0.07&2 \\
 & 50255$^{+}$ &  $-$1.27&9&$-$0.38&5 & & & 52007 &$-$0.09&3&0.06&2 \\
 & 50351  &   2.2& 3&$-$0.64&9 & & 1642$-$03 & 47156 & 2.15&7&1.93&4 \\\\
 & 50414  & $-$1.07& 9&0.23&4 & & & 47740& 0.45&3&$-$1.28&3 \\
 & 50646 &  $-$2.6&3&$-$0.88&6 & & & 48164$^{+}$ &  $-$2.81&3&1.79&6 \\
 & 50824  &  $-$1.3 &3&0.41&8 & & & 48939  &   0.33&2&$-$0.51&1\\
 & 50970$^{+}$ &  4.5&8&1.7&5 & & & 49358  &   1.16& 9 &0.70&6 \\
 & 51167  &   1.6&2&$-$0.24&8 & & & 49646  &  $-$0.42&2&$-$0.85&2\\
 & 51950 &  $-$1.8&4&$-$0.5&1 & & & 50280  &  $-$0.59&3&1.24&3\\
 & 52073  &  $-$2.1&4&0.39&6& & & 50650$^{+}$ & 1.9&2&1.48&10 \\
 & 52264$^{+}$ &   1.9 &2&0.67&4 & & & 50978  &  $-$0.22&3&$-$1.59&2 \\
1356$-$60 & 47855 & 0.21&4&$-$0.12&1& & & 51173&$-$0.8 & 2 &$-$1.3&2 \\
 & 48680  & $-$1.9&5&$-$0.9 &2 & & & 52183$^{+}$ & $-$1.85&8&1.72&2 \\\\
 & 48847& 0.45&7&$-$0.033&8 & & 1706$-$16&48131$^{+}$&$-$3.5&8&2.7&5\\
 & 49388  &  0.38&6&$-$0.06&2 & & & 48535 &  $-$0.89& 9 & $-$1.7&4 \\
 & 49685  &   0.65&4&0.047&3 & & & 48615  & 0.56 & 5 & 3.3 &3 \\
 & 49713  &  $-$0.53&8&$-$0.04&1 & & & 49101$^{+}$ & $-$3.6 &5& $-$3.5&6 \\
 & 50101$^{+}$ & $-$1.68&9&$-$0.85&1 & & & 49257  &  1.15&8&$-$0.65&3 \\
 & 50356  &  0.022&4&0.049&8 & & & 50541  &  1.0&2&0.31&5 \\
 & 50757  &  0.26& 6 &$-$0.056&6& & &51002& 0.30&6&$-$0.22&1\\
 & 51150  &   0.8& 2 &0.85&3& & & 51830 &$-$0.48&4&$-$0.36&6 \\
 & 52009 &$-$0.12&4&$-$0.07&1& & 1727$-$47&46383&$-$1.5 &5 &0.06&2 \\
1358$-$63&47725$^{+}$&$-$1.8& 3&$-$2.38 & 9& & &46913&$-$1.2&4&0.034&8
\\\hline
\multicolumn{13}{l}{$^{a}$ The formal standard error and refers to the
 least significant digit}   \\
\multicolumn{13}{l}{$^{+}$ Indicates epochs of events whose jump parameters
were better resolved by the manual technique}
\\\hline
\end{tabular}
\end{minipage}
\end{table*}

\begin{table*}
\centering
\begin{minipage}{140mm}
\contcaption{}
\begin{tabular}{@{}llrrrrcllrrrr}
\hline
\multicolumn{1}{l}{Pulsars} & \multicolumn{1}{l}{Epoch} & 
\multicolumn{1}{r}{$\Delta{\nu}/\nu$} &
\multicolumn{1}{r}{ERR$^{a}$} &
\multicolumn{1}{r}{$\Delta{\dot{\nu}}/\dot{\nu}$} 
&  \multicolumn{1}{r}{ERR$^{a}$} &  & \multicolumn{1}{l}{Pulsars} &
\multicolumn{1}{l}{Epoch} & \multicolumn{1}{r}{$\Delta{\nu}/\nu$} &
\multicolumn{1}{r}{ERR$^{a}$} &
\multicolumn{1}{r}{$\Delta{\dot{\nu}}/\dot{\nu}$}
&  \multicolumn{1}{r}{ERR$^{a}$} \\
(PSR B) &  (MJD) & ($10^{-9}$)  &   & ($10^{-3}$) &  &  & (PSR B) &  (MJD) &
($10^{-9}$) &   & ($10^{-3}$) &  \\
(1)  & (2)  & (3) & (4) & (5) & (6) &  & (1)  & (2)  & (3) & (4) & (5) & (6)
\\\hline
 & 47123& 0.5 & 1 &$-$0.017&2& & & 46979  &   0.92&9&0.42&8 \\
 & 47260$^{+}$&9.3 & 3&$-$1.95&6& & &47132& 1.2&6&$-$0.48&9 \\
 & 47499 &$-$0.33&4&$-$0.07&2& & & 47322$^{+}$ & 3.5 &6& 0.75&8\\
 & 47889  &   0.6 & 2 &0.50&5& & & 47764  &   2.4&7&0.3&1 \\
 & 48540  &  $-$0.26&8&0.08&1& & & 48140$^{+}$ & 1.9&5& 0.25&5\\
 & 49200$^{+}$&13.6 &8&$-$2.54 &2 & & &48210 &$-$2.07&8&0.5&2\\
 & 49817 &   0.84&7&$-$0.12&1 & & &48451  &   0.6&2&0.08&2 \\
 & 50088  &$-$1.13&6&$-$1.04&2& & & 48580 & $-$7.8&2&$-$0.89&9 \\
 & 50388&0.97&7&0.033&1 & & & 48747  &  $-$0.9&4&$-$0.16 &4\\
 & 50662& 1.3 &3&0.38&8 & & &49132&$-$1.7&6&$-$0.7& 3 \\\\
 & 52108 & 0.7&2&0.05&1 & & & 49279  &  $-$4.7&9&0.6& 2 \\
 & 52300 & 3.76 & 8 & $-$0.58& 6 & & & 49876& 3.6&2&$-$0.38&2 \\
1749$-$28 & 46805 & $-$0.50&4&0.12&1& & & 50150$^{+}$& 4.9 & 3& $-$2.35& 7 \\
  & 46990$^{+}$& 5.03 & 7 & $-$1.6 & 2& & & 50495  &  $-$1.2&2&0.18&6 \\
  & 47315 &  0.58&2&0.23&2 & & & 50585  & $-$0.8&2&$-$0.28&3 \\
  & 47660 & $-$0.31&2&$-$0.13&1 & & & 50997  &$-$0.98&4&$-$0.68&9\\
  & 48098$^{+}$ & 3.5 &8&0.79&3 & & & 51905$^{+}$ & 32 &3& 1.68&8 \\
  & 48485 &  0.48&6&$-$0.42&4 & & & 52041  & $-$0.83&4&$-$0.74&9 \\
  & 48701$^{+}$ & $-$1.3 & 2&0.89 &6 & & & 52110 & 0.8&3 &1.92&8  \\
  & 48806 & $-$0.96&6&0.36&5 & &  & 52273  &  12.6&6  & $-$2.22&5 \\\\
  & 48916$^{+}$ &2.3&5&0.62&9 & &1929+10&47391&$-$0.027&6&$-$0.067&8\\
  & 49173 & $-$0.32&6&$-$0.18&5 & & &47670$^{+}$&$-$0.62&4 &$-$0.21 &8\\
  & 49756$^{+}$ &  0.8&2&$-$0.6 & 1& & & 48360$^{+}$ & 0.74&7&$-$0.39&2\\
  & 49964 &  0.30&5&0.57&3 & & & 48446  &  $-$0.28 & 2&0.06&1 \\
  & 50260 &  1.09&6&0.41&3 & & & 49064  &  $-$0.23 & 4&$-$0.16&2 \\
  & 50356 & $-$0.33&5&$-$0.39&4 & & & 49205  &  $-$0.27 & 5&0.15&2\\
  & 50408$^{+}$ &  0.56&6&$-$1.12&8 & & &49542& $-$0.16 & 5&0.18&3\\
  & 50583 & $-$0.31&6&0.13&4 & & &49710&$-$0.013 & 2&$-$0.012&2 \\
  & 50727 &  0.68&9&$-$0.15&2& & & 50182  &  $-$0.05 & 1&0.33&4 \\
  & 51108$^{+}$ & 1.2 &2&$-$0.65&7 & & &50266&$-$0.07 & 2&$-$0.38&3\\\\
  & 51250 &  0.65&6&$-$0.21&1 & & & 50773  &   0.09&2 & 0.14&2\\
  & 52075$^{+}$ & $-$0.76&4&0.19&1 & & &51198& 0.19&3 & 0.10&2 \\
1822$-$09&46838& $-$2.2&5&$-$0.17&6 & & &52081&0.5&1& $-$0.25&3\\\hline
\multicolumn{13}{l}{$^{a}$ The formal standard error and refers to the
 least significant digit}   \\
\multicolumn{13}{l}{$^{+}$ Indicates epochs of events whose jump parameters
were better resolved by the manual technique.}
\\\hline
\end{tabular}
\end{minipage}
\end{table*}

\begin{table*}
\centering
\begin{minipage}{140mm}
\caption{Parameters of the 33 microglitches, whose amplitudes are 
significant in only $\nu$ or $\dot{\nu}$, observed using the automated 
search technique}
\begin{tabular}{@{}llrrrrcllrrrr}
\hline
\multicolumn{1}{l}{Pulsars} & \multicolumn{1}{l}{Epoch} & 
\multicolumn{1}{r}{$\Delta{\nu}/\nu$} &
\multicolumn{1}{r}{ERR$^{a}$} &
\multicolumn{1}{r}{$\Delta{\dot{\nu}}/\dot{\nu}$} 
&  \multicolumn{1}{r}{ERR$^{a}$} &  & \multicolumn{1}{l}{Pulsars} &
\multicolumn{1}{l}{Epoch} & \multicolumn{1}{r}{$\Delta{\nu}/\nu$} &
\multicolumn{1}{r}{ERR$^{a}$} &
\multicolumn{1}{r}{$\Delta{\dot{\nu}}/\dot{\nu}$}
&  \multicolumn{1}{r}{ERR$^{a}$} \\
(PSR B) &  (MJD) & ($10^{-9}$)  &   & ($10^{-3}$) &  &  & (PSR B) &  (MJD) &
($10^{-9}$) &   & ($10^{-3}$) &  \\
(1)  & (2)  & (3) & (4) & (5) & (6) &  & (1)  & (2)  & (3) & (4) & (5) & (6)
\\\hline
0835$-$41&47437& $-$ &$-$ & $-$0.099 & 3& & & 50607 & $-$1.53 & 5 & $-$ & $-$ \\ 
 & 50961&$-$0.25 & 1 & $-$ & $-$& &1240$-$64 & 47423 & 0.26 & 2 & $-$&$-$ \\	
 & 51201 & $-$0.36 & 3 & $-$ & $-$ & &  & 48712 & $-$ & $-$ & $-$0.18  & 5 \\
0959$-$54 & 47570 & $-$0.95&4& $-$ &$-$& & & 49076 &$-$&$-$& $-$0.65 & 2\\ 	
 & 47983 & $-$2.37&4&$-$&$-$& &1323$-$62 & 49688 & $-$ & $-$ &  0.18&3 \\ 
 & 48017 & 1.55&5& $-$& $-$ & &1358$-$63 & 49733 & $-$ & $-$ &$-$0.47 & 2  \\ 	
 & 49076 & $-$ & $-$& $-$0.65&2 & & & 48555 & $-$ & $-$ & 0.69 & 3  \\
 & 49511 & $-$1.8 & 3 &$-$ & $-$ & & & 49767 & $-$ & $-$ & 0.49 & 6  \\
 & 50032 & 1.82& 6&$-$& $-$& & & 50251 & $-$ & $-$ & $-$0.69  & 8  \\	
 & 50455 & 1.29 &4 &$-$ &$-$& &1557$-$50 & 47599 & 1.29 & 9 & $-$ & $-$\\\\
 & 49711 & $-$ & $-$ & 0.25 & 6 & & & 49006 & $-$0.29 & 6 & $-$ & $-$ \\
 & 50696 & $-$ & $-$ &0.5 & 1& &1749$-$28 & 50395 & $-$& $-$ & 0.79& 5\\  
 & 50983 & 0.85&3&$-$ & $-$& & 1822$-$09 & 47960 & $-$&$-$&0.12&2 \\ 	
 & 51899 & 3.85  & 5 & $-$ & $-$ & & & 48797 & $-$ & $-$ & $-$0.039 & 6 \\ 	
 & 52210 & $-$2.2 &  2 & $-$ & $-$ & & & 50098 & $-$5.7  & 7  & $-$ & $-$ \\
1706$-$16 & 49880 & $-$0.86 & 8&$-$ & $-$ & & & 51908&$-$2.5&3&$-$ & $-$ \\ 	
1727$-$47 & 48886 & 0.35 & 5 & $-$ & $-$ & & & & & &  & 
\\\hline
\multicolumn{13}{l}{$^{a}$ The formal standard error and refers to the
 least significant digit}.\\
\\\hline
\end{tabular}
\end{minipage}
\end{table*}

\section{DISCUSSIONS}
\subsection{Macroglitches among the HartRAO pulsars}
Twenty-five of the twenty-six pulsars on routine
monitoring at HartRAO are aged more than $10^5$ years, having an average of
$10^{6.17}$ years. None of the twenty-five has shown any fractional
frequency jump larger than $\sim 10^{-8}$ within the seventeen years of
observation. This probably suggests that the interval for large macroglitches in 
these ``middle-aged'' (characteristic age $\sim 10^5\ - \ 10^7$ years) pulsars,
if they do occur at all, could be more than the total observation span of $\sim$ 
425 years.

The only one in the sample that has shown a large jump,
PSR 1727$-$47, is younger than $10^5$ years. Within this period, the pulsar
had two  large macroglitches of amplitudes $\Delta{\nu /} \nu = 1.37 
\times 10^{-7}$ and $1.25\times 10^{-7}$ at MJD 49382 and 52476,
respectively. There was also a smaller one, though larger than the jumps
reported here, at MJD 50718 with $\Delta{\nu /} \nu = 3.2 \times
10^{-8}$. Very many other microglitches have been reported in this paper for
this pulsar. 

{\bf This is in agreement with earlier suggestions that the oldest pulsars, like
the millisecond pulsars are  free from macroglitches. In a
statistical study of 48 macroglitches from 18 pulsars, Lyne et al.\ (2000)
found that the glitch activity decreases linearly with decreasing rate of
slowdown. Also, Urama \& Okeke (1999) reported that smaller values of
$\Delta{\nu}/\nu$ and $\Delta{\dot{\nu}}/\dot{\nu}$ are seen to be more
common in pulsars older than 10$^{5}$ years and that no glitch has been
observed in conventional radio pulsars older than 10$^{7}$ years.} 

\subsection{Microglitches and timing noise events}
The observed timing activity of 20 pulsars are found to be at levels (with 
$\sigma_{R} / \sigma_{W} > 5$) considered significant for meaningful
microglitch analyses. For the remainder of 6 pulsars, the observed timing activity 
over $\sim$ 13 yr period is almost indistinguishable from measurement uncertainty. 
The intrinsic scatter in the BTOAs of these objects have amplitudes in excess of 
20 ms (see Chukwude 2007) which could easily and effectively swamp any low level 
timing activity in the pulsars. This anomalously large scatter in the data could be 
attributed to HartRAO local observing system parameters: high observing frequencies 
(13 and 18 cm) and the narrow receiver band of 10 MHz. These parameters
could combine to further reduce the precision of measurements of the pulse
times of arrival. This would ultimately degrade the BTOAs of pulsars which
intrinsically exhibit  low level of timing activity. Another
possibility is that some of the pulsars are intrinsically faint objects.
For instance, previous independent analyses of, presumably, higher precision, 
timing data on some of these pulsars (1133+16, 1426$-$66, 1451$-$68 and 
2045$-$16) do not reveal any appreciable level of timing activity (DA95; CD85). 

For the sub-sample of 20 pulsars extensively used in current analyses, the 
results are both phenomenal and unprecedented in the history of microglitch
observations in radio pulsars. The automatic technique identified a total
of 266 jumps that are simultaneously significant in both pulsar rotation 
frequency and spin-down rate. Only 71 (or $<$ 30\%) of these microglitches
were equally resolved by the traditional manual method.  In addition, the 
automated technique observed 19 and 14
significant jumps in only $\nu$ and $\dot{\nu}$, respectively. The fractional jumps
in $\nu$ and $\dot{\nu}$ ($\Delta{\nu}/\nu$ and $\Delta{\dot{\nu}}/\dot{\nu}$, 
respectively) have all possible combinations of signs. For the 266 microglitch 
events, we found ($\Delta{\nu}/\nu$, $\Delta{\dot{\nu}}/\dot{\nu}$) = (+,
+), (+, $-$), ($-$, $-$) and ($-$, +) as the signatures of 69, 68, 66 and
63 events, respectively. It is worthy of note that the distribution of
these events is not statistically different from random. Our results also reveal large dispersions, $\sim$ 4 
orders of magnitude, in observed sizes of the jumps in both spin parameters.
Strikingly, the observed fractional jumps in $\nu$  and $\dot{\nu}$
have sizes whose magnitudes are in the range of $\sim$ 2$\times10^{-11}$ 
$-$ 2$\times10^{-8}$ and 5$\times10^{-6}$ $-$ 4$\times10^{-3}$, respectively. 
Typical median values for $|\Delta{\nu}/\nu|$ and $|\Delta{\dot{\nu}}/\dot{\nu}|$
are $\sim$ $0.8\times10^{-9}$ and $0.4\times10^{-2}$, respectively.

Arguably, we have presented what is as yet the most in-depth and successful 
observations of microglitch phenomenon in  slow radio pulsars. 
Nonetheless, there had been some attempts in the past to observe these small
amplitude discrete events in pulsar rotation rates. Cordes \& Downs (1985) 
carried out extensive analyses of the BTOAs of 17 pulsars obtained at Jet
Propulsion Laboratory (JPL) between 1968 and 1982. In the framework of the
manual technique, the authors identified a total of 74 candidate events 
in $P$ and $\dot{P}$. However, only 43 of these events were found to be
significant $-$ 10 simultaneously in $P$ and $\dot{P}$, while 17 and 16 
were separately significant in $P$ and $\dot{P}$, respectively. Similar 
analysis, based on a sample of 26 pulsars observed at Mt. Pleasant Observatory  
between 1987 and 1991 (D'Alessandro et al.\ 1995) yielded about 35 
significant microjump events from about 70 candidate events. A breakdown 
of the events shows that while 5 events were noticeable in both $\nu$ and 
$\dot{\nu}$, 20 and 10
were significant only in $\nu$ and $\dot{\nu}$, respectively. In comparison,
current analyses identified $\sim$ 300 microjumps from just 20 pulsars,
which is about a factor of 4 in excess of all the hitherto known radio pulsar
microglitches.
 
As a check on the reliability of the
methods and the reality of the results, it might be necessary to compare it
with those of previous studies. It is noteworthy that the HartRAO and
Tasmanian timing data on 16 pulsars overlap each other, up to 1991.
Remarkably, over 90\% of the microglitch events reported for these objects
by DA95 also show up in current analyses and the results (epochs and
signatures of the events) are generally in good agreements. However, in terms 
of the jump sizes, our results differ considerably with those published by 
DA95, nonetheless they are still of the same order of magnitude.
Irrespective of sign, we find that the amplitudes of the current microjumps 
are, in most cases, considerably larger than those reported by DA95,
suggesting a better resolution of the microevents. This is particularly true
for $\Delta \dot{\nu}$, where our measurements could be up to a factor of 4
larger than those observed by DA95. As a consequence, most of the candidate 
microglitch events found to be insignificant in DA95 were observed to be 
significant in current analysis.         

The unprecedented improvements in both the number and amplitudes of pulsar
microglitches could be attributed to some unique features of the current timing
data: long time coverage ($\sim$ 16 yrs) and shorter sampling intervals.
Shorter data sampling intervals would naturally lead to better time resolutions 
of events. Expectedly, this would reduce the uncertainty in event epoch, 
resulting in significantly improved estimates of jump sizes. Glitch analyses 
(e.g.\ Flanagan 1990) have demonstrated how the resolution of jump parameters 
strongly depended on the accuracy of the epochs of the events. The rise times 
of the events ($\Delta{t}$) for current 
analysis lie between 0.05 and 40 d, with a median value of $\sim$ 10 d. This
implies that the epochs of the microglitches reported in this paper are
improved by a factor of, at least, 4 over those of JPL and Tasmanian data
sets, where  $\Delta{t}$ are in the range of $\sim$ 6 $-$ 154 d and 13 $-$ 
255 d, respectively. This could, almost certainly, lead to better
resolution of event sizes. It is noteworthy that our techniques identified
most of the small macroglitches reported previously from independent analyses
timing data on the pulsars B0740$-$28 and B1240$-$64 (Janssen \& Stappers
2006; D'Alessandro \& McCulloch 1997). Again, our results on the jump parameters
(epochs, amplitudes and signatures) are remarkably in good agreement with
previous results. However, current methods were able to identify almost all 
the jumps in $\dot{\nu}$, which previous analyses could not resolve owing to
large uncertainties in event epochs. Hence, the seemingly high sensitivity of the 
current analysis could be attributed to the dramatic improvement in precision 
of the events' epochs in HartRAO data set.

A key challenge in microglitch observation analyses has remained the
extraordinary difficulty in distinguishing between real and spurious events. 
Cordes \& Downs (1985) suggested that microjumps could be identified by 
demostrating that they are too large to be mere fluctuations produced by 
accumulation of many, much smaller events of random walk origin.
Previous studies (DA95; Cordes et al.\ 1988; CD85) have relied largely on 
the signficance tests, using timing noise strength parameters, to isolate 
significant events. However, Cordes \& Downs (1985) have highlighted the
strong dependence of the current method of testing the significance of
microglitch events on the rise time of the event $\Delta{t}$. For instance, 
the standard deviations could possibly be 
over-estimated or under-estimated for very small or very large $\Delta{t}$, 
respectively. This could cause leakage of spurious events with large
amplitudes or rejection of extremely small-amplitude discrete events. 
The former scenario might be the case in previous analyses (DA95; CD85), 
while current data appear to favour the later  scenario. With a median event 
rise time ($\Delta{t}$) of $\sim 10$ d, it is feared that both 
$\sigma_{\Delta{\nu}}$ and $\sigma_{\Delta{\dot{\nu}}}$ could have been 
significantly under-estimated 
in some cases, causing significant leakage of spurious events into the 
sample of real discrete events (e.g.\ Chukwude 2002). However, we are 
optimistic that the 2-$\sigma$ standard formal error condition imposed on 
the size identified microglitches is stringent enough to filter out all or, at 
least, most spurious events. 

\section{SUMMARY}
The phenomenon of microglitches in slow radio pulsars have been probed deeply 
using the manual and automatic search techniques. An extensive analysis of 
a sub-sample of 20 pulsars, whose timing data span $\sim$ 16 yrs in time and
have a median sampling frequency of $\sim$ 0.1 d$^{-1}$, yielded a phenomenal
299 microglitch events in pulse rotation frequency and its first derivative.
This translates to more than a factor of 3 increase in the statistics of 
the observed microglitch events. The jumps show no preferred signs and have 
amplitudes that span about 3 orders of magnitude in both parameters, irrespective 
of signs. Current results have, among other things, demonstrated the prevalence 
of microglitch phenomenon in slow radio pulsars. 
 
\section*{Acknowledgments}

This work was done, in part, when AEC was visiting the Abdus Salam International
Centre for Theoretical Physics, Trieste, Italy as a Junior Associate. He
is grateful to the Swedish International Development Agency (SIDA) for
supporting his visit to ICTP with a travel grant. The authors wish to
acknowledge the Director of HartRAO and Dr. C. S. Flanagan for giving them
access to the Observatory pulsar data.

\bsp

\label{lastpage}

\end{document}